\documentclass[prd,twocolumn]{revtex4}
\usepackage{dcolumn}
\usepackage{multirow}
\usepackage{graphicx}
\usepackage{amssymb}
\usepackage{bm}
\usepackage{hyperref}
\usepackage{epstopdf}
\usepackage{color}
\usepackage{mathrsfs}
\usepackage{amsmath,amssymb,amsthm}
\usepackage{rotating}
\usepackage{sverb, longtable}
\usepackage{subfigure}
\usepackage{setspace}

\usepackage[graphicx]{realboxes}
\usepackage{adjustbox}
\begin{document}

\title{Constraining Cosmological Phase Transitions with Chinese Pulsar Timing Array Data Release 1}
\author{Deng Wang}

\email{cstar@nao.cas.cn}
\affiliation{Instituto de F\'{i}sica Corpuscular (CSIC-Universitat de Val\`{e}ncia), E-46980 Paterna, Spain \\
	National Astronomical Observatories, Chinese Academy of Sciences, Beijing, 100012, China}
\begin{abstract}
The Chinese Pulsar Timing Array (CPTA) collaboration has recently reported the observational evidence of a stochastic gravitational wave background. In light of the latest CPTA observation, we aim at exploring the ability of CPTA in probing new physics. Specifically, we constrain the first-order cosmological phase transitions with CPTA data, and find that the constraining result is slightly tighter than that of NANOGrav's 12.5-yr data but weaker than NANOGrav's 15-yr data. Considering the possible complexity of gravitational wave sources, we give the constraint on a mixed scenario of cosmological phase transitions and astrophysical supermassive binary black holes. Our analysis suggests that CPTA has a great potential to probe fundamental physics in the near future.

\end{abstract}
\maketitle

\section{Introduction}
Human beings have entered the multi-messenger era to probe the universe, since LIGO-VIRGO collaboration reports the first binary black hole coalescence GW150914 at high frequencies \cite{LIGOScientific:2016aoc}. In the near future, the space-based gravitational wave (GW) detectors such as LISA \cite{LISA:2017pwj} will help explore the low frequency GW sources such as massive binaries and supernovae. Moreover, pulsar timing arrays (PTA) \cite{Manchester:2013ndt,Dewdney2009} will detect the stochastic gravitational wave background (SGWB) at very low frequencies. PTA experiments detect the SGWB by the correlated deviations from time of arrivals (TOA) of radio pulses for a network of precisely timed millisecond pulsars within $\sim$ 1 kpc region from the earth. So far, there are three independent PTA groups searching for SGWB using their long-term accumulated TOA data: (i) North American Nanohertz Observatory for Gravitational Waves (NANOGrav) \cite{Brazier:2019mmu}, European Pulsar Timing Array (EPTA) \cite{Desvignes:2016yex} and Parkes Pulsar Timing Array (PPTA) \cite{Kerr:2020qdo}. The integration of these three groups forms the so-called International Pulsar Timing Array (IPTA) \cite{Perera:2019sca}.       

Most recently, it is very exciting that NANOGrav \cite{NANOGrav:2023gor}, EPTA \cite{Antoniadis:2023rey} and PPTA \cite{Reardon:2023gzh} have simultaneously reported substantial evidences of a stochastic common spectrum process at very low frequencies with a higher confidence level than their previous results  \cite{Brazier:2019mmu,Desvignes:2016yex,Kerr:2020qdo}. The correlations follow the Hellings-Downs pattern expected for the SGWB. In particular, the Chinese Pulsar Timing Array (CPTA) also reports the observation evidence of the SGWB using the data release one (DR1) from the Five-hundred-meter Aperture Spherical radio Telescope (FAST) \cite{Xu:2023wog}. The origin of an SGWB could be astrophysical or cosmological. In general, the astrophysical GW source is the merger of supermassive binary black holes (SMBBH), while the cosmological GW sources are complex such as early cosmological phase transitions \cite{NANOGrav:2023hvm,Kosowsky:1992rz, Caprini:2010xv, Nakai:2020oit, Li:2021qer,Fujikura:2023lkn,Bringmann:2023opz}, axion-like particles \cite{Ratzinger:2020koh,Wang:2022rjz,Blasi:2022ayo}, cosmic strings \cite{Siemens:2006yp, Blanco-Pillado:2017rnf,Ellis:2020ena,Blasi:2020mfx}, domain wall decaying \cite{Wang:2022rjz,Hiramatsu:2013qaa,Ferreira:2022zzo}, inflation \cite{Vagnozzi:2023lwo}, large primordial curvature perturbations \cite{Kohri:2018awv}, primordial magnetic field \cite{RoperPol:2022iel} and so on.  

If cosmological phase transitions are of first order and last for a sufficiently long duration, they can serve as a potential source of the SGWB \cite{Hogan:1983ixn,Witten:1984rs,Hogan:1986qda,Turner:1990rc}. In this study, we focus on the very low frequency GWs produced by the first-order cosmological phase transitions. There are at least two phase transitions predicted by the standard model of particle physics in early universe, i.e., electroweak phase transitions at $T_\star\sim 100$ GeV and QCD phase transitions at $T_\star\sim 0.1$ GeV. The former is related to the electroweak symmetry breaking, while the latter is related to the chiral symmetry breaking. Up to now, there have been many cosmological, astrophysical and laboratorial probes to investigate the nature of the so-called hidden sectors \cite{Strassler:2006im,Chacko:2004ky,Schwaller:2015tja,Battaglieri:2017aum}. With the rapid development of observational techniques and gradually accumulated data of millisecond pulsars, PTA observations are verified to have the ability to explore the dynamics of hidden sectors. Specifically, NANOGrav \cite{NANOGrav:2021flc} reported that the observations can be explained with a strong first-order phase transition occurring below the electroweak scale and that a first-order phase transition is highly degenerated with SMBBH mergers as GW sources. PPTA \cite{Xue:2021gyq} found that pulsar timing is very sensitive to low-temperature phase transition lying in the range $T_\star\sim1-100$ MeV and can be used for constraining QCD phase transitions. Other related works that use PTA observations to constrain phase transitions can be found in Refs.\cite{Ratzinger:2020koh, Moore:2021ibq}.

In light of the CPTA DR1 which consists of TOA measurements and pulsar timing models from 57 pulsars around the frequency of 14 nHz, we attempt to constrain the cosmological phase transitions. The data covers the time span between April 2019 and September 2022 and corresponding observations were conducted using FAST \cite{Xu:2023wog}. After numerical analysis, we find that the constraining result of CPTA DR1 is slightly tighter than that of NANOGrav's 12.5-yr data but weaker than NANOGrav's 15-yr data.

\section{First order phase transitions}
First order cosmological phase transitions happen through the locally tunneling of a field when there is a barrier between a false minimum and a true minimum of a potential. In the early universe, this kind of phase transitions are conducted by the nucleation of true vacuum bubbles, which expand over time in the background plasma. The ultra low frequency GWs can be generated by collisions of a great number of bubbles and interacting bubble walls and background plasma.

\begin{table*}[htbp]
	\renewcommand\arraystretch{2}
	\caption{Parameters and formula for the GW energy spectrum from the cosmological phase transitions.}
	\setlength{\tabcolsep}{5mm}{
		{\begin{tabular}{@{}cccc@{}} \toprule
				Parameters      &Bubble walls      &Sound waves           &Turbulence                               \\ \colrule
				$\kappa$       &$\kappa_\phi$  &  $\kappa_{\mathrm{sw}}$    &$0.1\times\kappa_{\mathrm{sw}}$          \\  
				$p$          &2  &2         &1.5                              \\
				$q$                  &2       &1        &1                          \\
				$\Delta(v_w)$ &$\frac{0.48v_w^3}{1+5.3v_w^2+5v_w^4}$   &$0.513v_w$    &$20.2v_w$    \\
				$\frac{f_\star}{\beta}$   &$\frac{0.35}{1+0.07v_w+0.69v_w^4}$   &$\frac{0.536}{v_w}$   &$\frac{1.63}{v_w}$   \\
				$S(x)$   &$\frac{(a+b)^c}{(ax^{b/c}+bx^{-a/c})^c}$              &$x^3\left(\frac{7}{4+3x^2}\right)^{3.5}$          &$\frac{x^3}{(1+x)^{11/3}(1+8\pi xf^0_\star/\tilde{H}_\star)}$             \\
				\botrule
			\end{tabular}
			\label{t1}}}
\end{table*}

There are three main low frequency GW sources in the model of phase transitions \cite{Weir:2017wfa, Breitbach:2018ddu} including the bubble collisions \cite{Caprini:2007xq, Huber:2008hg}, collisions of sound wave originated from bubbles expansion \cite{Hindmarsh:2013xza,Hindmarsh:2016lnk}, and the magnetohydrodynamics turbulence \cite{Caprini:2009yp} from bubbles expansion and sound wave collisions. Therefore, the total GW energy spectrum is written as $\Omega_{\mathrm{GW}}(f)=\Omega_{\mathrm{bub}}(f)+\Omega_{\mathrm{sw}}(f)+\Omega_{\mathrm{tur}}(f)$, where $f$ denotes the GW frequency. By adopting the standard 4-parameter model, the GW energy spectrum can be shown as \cite{Caprini:2009yp, Jinno:2016vai, Hindmarsh:2017gnf}    
\begin{equation}
\Omega_{\mathrm{GW}}(f)h^2=\mathcal{F}\Delta(v_w)S\left(\frac{f}{f^0_\star}\right)\left(\frac{H_\star}{\beta}\right)^q\left(\frac{\kappa\alpha_\star}{1+\alpha_\star}\right)^p,  \label{1}
\end{equation}
where $\mathcal{F}=7.69\times10^{-5}g_\star^{-\frac{1}{3}}$ denotes the redshift of GW energy density, $g_\star$ the number of relativistic degree of freedom, $\Delta(v_w)$ a normalization factor depending on the bubble wall velocity $v_w$, $\kappa$ the efficiency factor, $\alpha_\star$ the strength of phase transitions that determines the amplitude of GW energy spectrum, $H_\star$ the Hubble parameter at the energy scale of phase transitions $T_\star$, $\beta$ the inverse duration of phase transitions, and the function $S(f/f_\star^0)$ depicts the spectral shape, where the present peak frequency $f_\star^0$ is expressed as phase transitions 
\begin{equation}
f_\star^0\simeq 1.13\times10^{-10}\left(\frac{f_\star}{H_\star}\right)\left(\frac{T_\star}{\mathrm{MeV}}\right)\left(\frac{g_\star}{10}\right)^{\frac{1}{6}}\mathrm{Hz}, \label{2}
\end{equation} 
and $S(x)$ from bubble collisions is characterized by free parameters $a$, $b$, $c$ \cite{Jinno:2016vai} and has the following form
\begin{equation}
S(x) = \frac{(a+b)^c}{(ax^{b/c}+bx^{-a/c})^c}.
\end{equation}

The values of normalization factor, efficiency factor, peak frequency at emission $f_\star$, spectral shape, and two exponents $p$ and $q$ are shown in Tab.\ref{t1}. $v_w$ and $\kappa$ are associated with $\alpha_\star$ and the dimensionless friction parameter $\eta$ \cite{Espinosa:2010hh}.

\begin{figure*}
	\centering
	\includegraphics[scale=0.5]{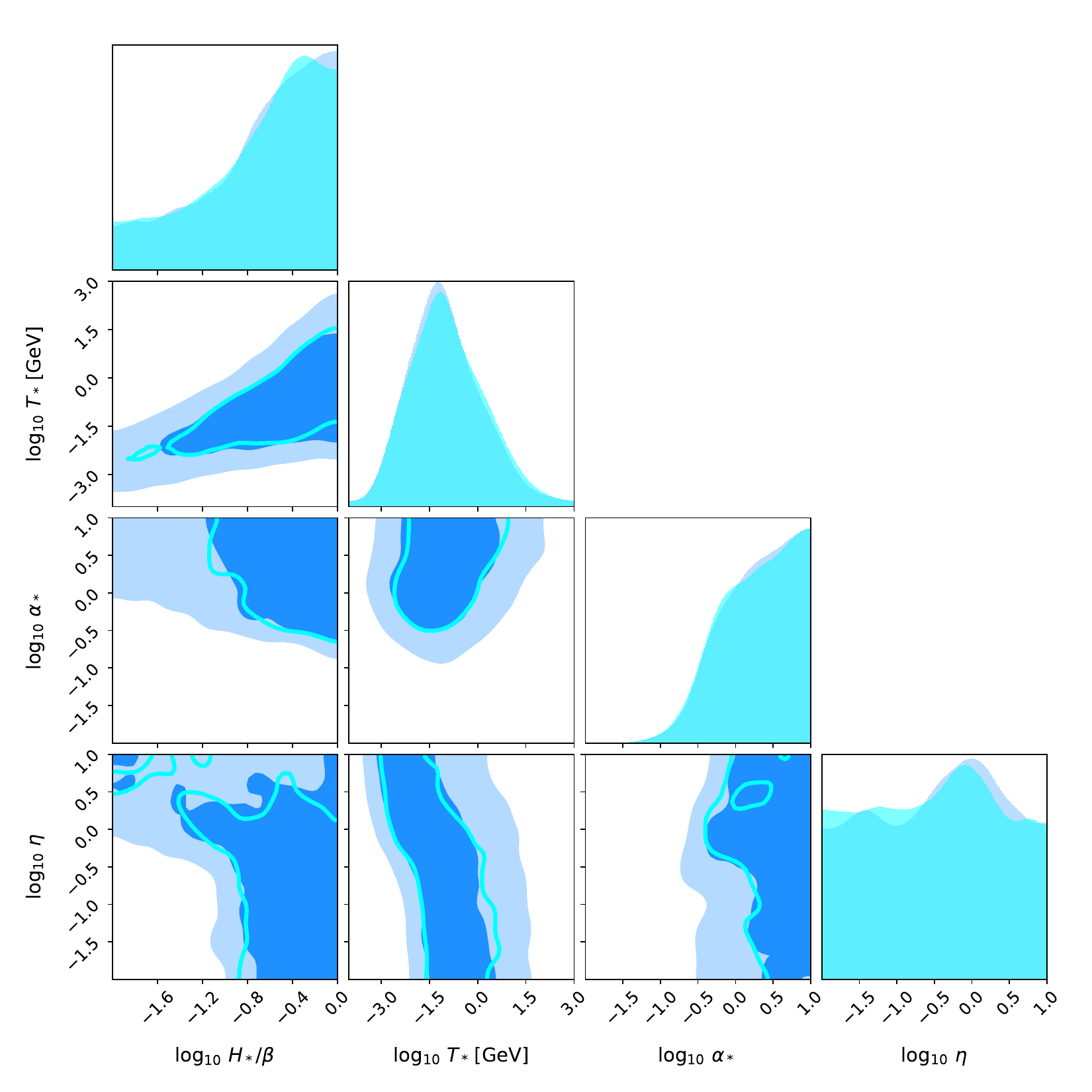}
	\caption{The marginalized posterior distributions of free parameters in the PTO model. The cyan contours represent the marginalized posterior distributions including the integrated constraint from CMB, BBN and astrometry. We use the semi-analytic approach and take the bubble spectra parameters $a=1$, $b=2.61$ and $c=1.5$.}\label{f1}
\end{figure*}

\begin{figure*}
	\centering
	\includegraphics[scale=0.6]{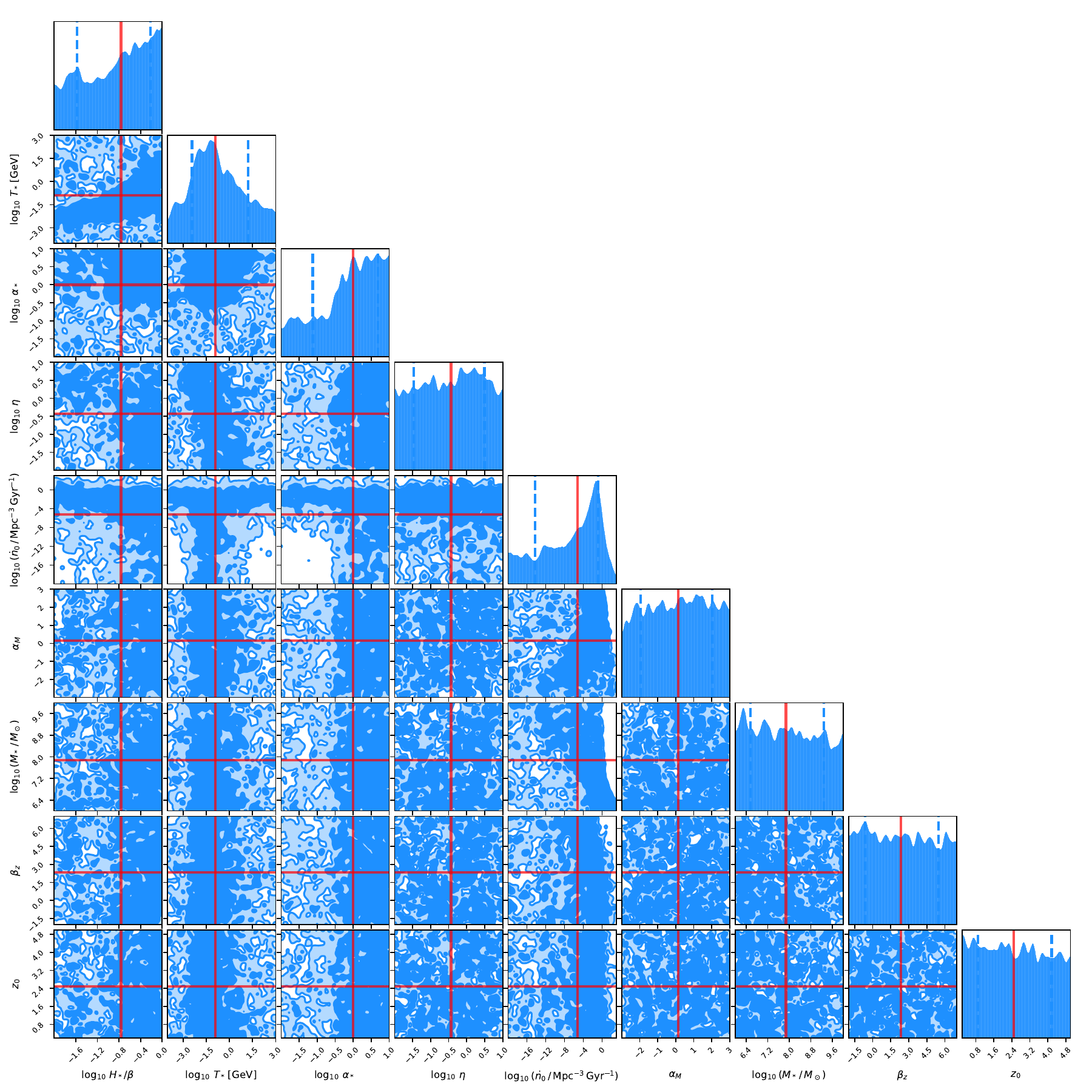}
	\caption{The marginalized posterior distributions of nine free parameters in the PTBBH model. The solid (red) and dashed (blue) lines are the mean values and $1\,\sigma$ bounds, respectively.}\label{f2}
\end{figure*}

\section{Supermassive binary black holes}

We introduce the details of astrophysical SMBBH model. Specifically, we take the binary black holes (BBHs) model which assumes the orbits of binaries are circular and the evolution of this system is only originated from GW emission \cite{Middleton:2015oda,Middleton:2020asl}. A population of BBHs are characterized by their mass function that represents the number density $n$ of BBHs per unit redshift $z$ per unit logarithmic chirp mass $\mathrm{log}\,\mathcal{M}$. The squared characteristic strain of the BBH model is shown as 

\begin{widetext}
	\begin{equation}
	h_c^2(f)=\frac{4G^{\frac{5}{3}}}{3\pi^{\frac{1}{3}}c^2}f^{-\frac{4}{3}}\int \mathrm{d}z(1+z)^{-\frac{1}{3}}\int\mathrm{d}(\mathrm{log}\,\mathcal{M})\mathcal{M}^{\frac{5}{3}}\frac{\mathrm{d}^2n}{\mathrm{d}z\,\mathrm{d}(\mathrm{log}\,\mathcal{M})}, \label{1}
	\end{equation}
\end{widetext}
where $G$ and $c$ denote the gravitational constant and speed of light, respectively, and $\mathcal{M}=(m_1m_2)^{\frac{3}{5}}/(m_1+m_2)^{\frac{1}{5}}$ for a binary with independent masses $m_1$ and $m_2$. The BBH mass function is shown as
\begin{widetext}
	\begin{equation}
	\frac{\mathrm{d}^2n}{\mathrm{d}z\,\mathrm{d}(\mathrm{log}\,\mathcal{M})}=\dot{n}_0\frac{\mathrm{d}t_\mathrm{R}}{\mathrm{d}z}\left[(1+z)^{\beta_z}\mathrm{exp}\left(-\frac{z}{z_0}\right)\right]\left[\left(\frac{\mathcal{M}}{10^7M_\odot}\right)^{-\alpha_\mathcal{M}}\mathrm{exp}\left(-\frac{\mathcal{M}}{\mathcal{M_\star}}\right)\right], \label{2}
	\end{equation}
\end{widetext}
where $\dot{n}_0$ is the merger rate density of inspiralling BBHs, $t_\mathrm{R}$ denotes time in the source frame \cite{Phinney:2001di}, $\beta_z$ and $z_0$ describe the redshift evolution of the BBH population, and $\alpha_\mathcal{M}$ and $\mathcal{M_\star}$ characterize the shape of the BBH mass function.

\begin{table*}
	\renewcommand\arraystretch{1.5}
	\caption{The $1\,\sigma$ (68\%) confidence ranges of free parameters for the PTO and PTBBH models from CPTA DR1.}
	\setlength{\tabcolsep}{1mm}{
		{\begin{tabular}{@{}cccccccccc@{}} \toprule
				Parameters      &$\mathrm{log}_{10}H_\star/\beta$      &$\mathrm{log}_{10}T_\star$           & $\mathrm{log}_{10}\alpha_\star$                 &$\mathrm{log}_{10}\eta$           &$\mathrm{log}_{10}\dot{n_0}$ &$\alpha_M$ &$\mathrm{log}_{10}(M_\star/M_\odot)$   &$\beta_z$ &$z_0$                \\ \colrule
				PTO       &$-0.57_{-0.72}^{+0.40}$  &  $-1.12_{-0.97}^{+1.25}$    &$0.334_{-0.567}^{+0.462}$   &$-0.39_{-1.06}^{+0.87}$    &---    &---    &---     &---    &---      \\  
				PTBBH           &$-0.76_{-0.81}^{+0.55}$  &$-0.91_{-1.51}^{+2.13}$          &$-0.004_{-1.118}^{+0.689}$         &$-0.43_{-1.04}^{+0.93}$     &$-5.28_{-8.99}^{+4.41}$       &$0.14_{-2.08}^{+1.90}$   &$7.88_{-1.31}^{+1.40}$          &$2.34_{-2.98}^{+3.14}$    &$2.49_{-1.59}^{+1.68}$                    \\
				\botrule
			\end{tabular}
			\label{t2}}}
\end{table*}

\section{Methodology and results}
Ref.\cite{Xu:2023wog} gives the constraint on the common power law model in Fig.2. We sample the posterior distribution of two free parameters $\alpha$ and $A_c$, where $\alpha$ is spectral index and $A_c$ is the characteristic strain amplitude. When sampling many enough points, we can well approximate the posterior distribution of two model parameters. Then, we can easily derive the probability distribution of free GW spectrum around 14 nHz by using the formula $h_c(f)=A_{c}(f/1{\mathrm{yr}^{-1}})^{\alpha}$. Furthermore, we use the free GW spectrum to implement the constraints on the above mentioned PTO and PTBBH models.

To implement the constraints with CPTA DR1, we take two models into account. The first is the phase transition only model with four basic parameters $\{T_\star, \, \alpha_\star, \, H_\star/\beta, \, \eta\}$. Since the origin of SGWB may be a mixture of cosmological phase transitions and astrophysical SMBBH, we consider a mixed GW source model which allows the arbitrary overlapping contributions from phase transitions and SMBBH. In the following context, we shall call these two models as ``PTO'' and ``PTBBH'', respectively. To depict the background history of the PTBBH model, we take the Planck-2018 cosmology \cite{Planck:2018vyg}. 
Note that the PTBBH model contains nine free parameters by allowing an arbitrary combination of phase transitions and SMBBH GW sources.

Our numerical results are presented in Tab.\ref{t1} and the marginalized posterior distributions of free parameters in these two models are presented in Figs.\ref{f1}-\ref{f2}. Comparing with NANOGrav's 12.5-yr analysis \cite{NANOGrav:2021flc}, we find that the constraining power of CPTA DR1 is slightly stronger than that of NANOGrav's 12.5-yr data but weaker than NANOGrav's 15-yr data \cite{NANOGrav:2023gor}. Specifically, for the case of PTO, we obtain the constraint on the phase transition temperature $\mathrm{log}_{10}T_\star=-1.12_{-1.75}^{+2.59}$ at the $2\,\sigma$ confidence level. This reveals that the dark or QCD phase transitions occurring below 1.3 MeV are ruled out, and the permitted phase transition temperature range is [1.3 MeV, 29.5 GeV] (see Fig.\ref{f1}). Note that the $2\,\sigma$ constraint from NANOGrav's 12.5-yr analysis is [1 MeV, 100 GeV]. We find a weakly positive correlation between $\mathrm{log}_{10}T_\star$ and $\mathrm{log}_{10}H_\star/\beta$ indicating that phase transition energy scale increases with increasing bubble nucleation rate. Furthermore, compared to NANOGrav's 12.5-yr results \cite{NANOGrav:2021flc}, we obtain tighter lower bounds on phase transition duration $H_\star/\beta>0.014$ and strength $\alpha_\star>0.21$ and tighter upper bounds on the friction $\eta<8.12$ at the $2\,\sigma$ confidence level. Similar to NANOGrav 12.5-yr data, $\alpha_\star$ and $\eta$ are highly degenerate with left three parameters, respectively. It is worth noting that the inclusion of integrated constraint $\tilde{\Omega}_{\mathrm{GW}}<10^{-6}$ from CMB, BBN and astrometry hardly help reduce the PTO parameter space. This is because the amplitude of SGWB is always much smaller than $10^{-6}$. Hence, the integrated constraint can provide very limited constraining power.

It is interesting that, for the case of PTBBH, when considering simultaneously both astrophysical and cosmological GW sources, the phase transition parameter space are substantially enlarged but the total tendency is consistent with the PTO case (see Fig.\ref{f2}). We find that CPTA DR1 give the constraint on the BBH merger rate density $\dot{n}_0$, i.e., $\mathrm{log}\,(\dot{n}_0/\mathrm{Mpc^{-3}Gpc^{-1}})=-5.28^{+4.41}_{-8.99}$ for PTBBH. It is easy to see that CPTA DR1 just give loose constraints for the left four BBH parameters in PTBBH.  Moreover, in Tab.\ref{t1}, we observe that constraints on four phase transition parameters in PTBBH become very poor. This is natural because we increase the theoretical uncertainty and enlarge the parameter space. For example, the constraint $\mathrm{log}_{10}T_\star=-1.12_{-0.97}^{+1.25}$ in PTO is clearly stronger than $\mathrm{log}_{10}T_\star=-0.91_{-1.51}^{+2.13}$ in PTBBH at the $1\,\sigma$ confidence level.
More high precision pulsar timing datasets or integrated observations from different information channels are need to investigate these scenarios and break the parameter degeneracies.

\section{Discussions and conclusions}
Recently, the substantial evidence of an SGWB is successively reported by NANOGrav, EPTA and PPTA, while the CPTA also reports the observational evidence of an SGWB for the first time based on the data from largest single aperture radio telescope FAST. In this study, we are dedicated to investigate the ability of CPTA in probing early universe physics. To be more specific, we constrain the standard 4-parameter first-order cosmological phase transitions with CPTA DR1. Comparing the result with NANOGrav's analysis, we find that the constraining power of CPTA DR1 is slightly tighter than that of NANOGrav's 12.5-yr data but weaker than NANOGrav's 15-yr data. The allowed phase transition temperature range is [1.3 MeV, 29.5 GeV] at the $2\,\sigma$ confidence level. Furthermore, the inclusion of integrated constraint from CMB, BBN and astrometry hardly help compress the PTO model parameter space. In addition, we give the constraint on the mixed PTBBH model and obtain the BBH merger rate density $\mathrm{log}\,(\dot{n}_0/\mathrm{Mpc^{-3}Gpc^{-1}})=-5.28^{+4.41}_{-8.99}$ at the $1\,\sigma$ confidence level.

It is noteworthy that the time span of CPTA DR1 is from April 2019 to September 2022, which is much smaller the integration time of NANOGrav. It is exciting that its constraining power of cosmological phase transitions has been comparable to that of NANOGrav 12.5-yr data. As a consequence, in the near future, we can expect high-precision CPTA pulsar timing data to probe the SGWB and new physics based on FAST observations. 

\section{Acknowledgements}
DW thanks Liang Gao, Kejia Li and Heng Xu for helpful discussions.

\end{document}